# ON ANALYSIS AND GENERATION OF SOME BIOLOGICALLY IMPORTANT BOOLEAN FUNCTIONS

**CAMELLIA RAY, JAYANTA KUMAR DAS, PABITRA PAL CHOUDHURY**

**ABSTRACT:** Boolean networks are used to model biological networks such as gene regulatory networks. Often Boolean networks show very chaotic behaviour which is sensitive to any small perturbations. In order to reduce the chaotic behaviour and to attain stability in the gene regulatory network, nested Canalizing Functions (NCFs) are best suited. NCFs and its variants have a wide range of applications in systems biology. Previously, many works were done on the application of canalizing functions, but there were fewer methods to check if any arbitrary Boolean function is canalizing or not. In this paper, by using Karnaugh Map this problem is solved and also it has been shown that when the canalizing functions of $n$ variable is given, all the canalizing functions of $n+1$ variable could be generated by the method of concatenation. In this paper we have uniquely identified the number of NCFs having a particular Hamming Distance (H.D) generated by each variable $x_i$ as starting canalizing input. Partially NCFs of 4 variables has also been studied in this paper.

## 1. INTRODUCTION

Idea of canalization was given by Kauffman [1]. Canalizing function is a kind of boolean function in which output of the boolean function can be predicted by the input of at least one variable. For example if the boolean function $f = x_1 \cup x_2 \cup x_3$ is considered and if the input for any one of the variables is 1 output of the function will be always 1, so through input of only one variable the output of the function can be obtained. For non canalizing function like $f = x_1 \oplus x_2$ input for both the variables are needed to obtain the output of the function. [2][3][15] gives an idea and also formula for finding the upper bound, the number of canalizing functions and nested canalizing functions in $n$ variable boolean function. It does not give any method for identification of any arbitrary $n$ variable boolean function. This problem of identification of any arbitrary boolean function as canalizing function has been solved in this paper using K-Map. Identification of canalizing function by the help of Karnaugh Map avoided many arithmetic computations which were done for identification of canalizing function using semi tensor product [16]. Behaviour of biological systems can be reflected by canalizing functions [4]. It has been seen in [5] [6] [10] that canalizing function and its variants proved to be very useful for identification of gene regulatory network. Prediction of protein structures, their functions, stability and phase transition of boolean network with respect to canalizing function were done in [7][14]. Some ordered behaviour were observed in the boolean network when they were described by canalizing rules [8] [9]. The dynamics of the boolean network with regards to canalizing functions, their extension and characteristics over a finite field has been discussed in [11] [12] [13].

Karnaugh Map [19] is mainly used for simplification of boolean network [17]. In this paper an attempt has been made to detect if any arbitrary given function is canalizing or not. It has been observed that by the method of concatenation also all the canalizing functions in $n+1$ variables can be detected from the canalizing function of $n$ variables. Different properties of the canalizing function have also been described by the help of Karnaugh Map. Here one formula has been derived to find the number of nested canalizing functions which is generated uniquely having a particular hamming distance $j$ with starting canalizing input $x_i$. For 4 variables the number of partially nested canalizing functions having different depths was also calculated.

In section 2, some preliminaries on canalizing function, Karnaugh Map and theorem for identification of canalizing function has been included. In section 3, some properties of the canalizing functions, and their generation in $n+1$ variable from $n$ variable were explained. In section 4 and 5 some analysis on nested canalizing and partially nested canalizing function has been discussed.

## 2. PRELIMINARIES
### 2.1 Karnaugh Map(K-Map)
Karnaugh Map or K-Map as defined in [19] is mainly used for simplifying Boolean algebraic expression. Here pattern recognition is used to avoid the extensive calculations. Boolean results are transferred from a truth table to a two dimensional matrix where the cells are ordered in Gray Code and each cell position represents one combination of input conditions and each cell value represents corresponding output value. For three variable $(x_1, x_2, x_3)$ boolean functions the truth table is as shown in Table 1 and the K-Map of result $\sum(1,3,5,7)$ is given in Table 2

**Table 1**

| $X_1$ | $X_2$ | $X_3$ | $f$ |
|---|---|---|---|
| 0 | 0 | 0 | 0 |
| 0 | 0 | 1 | 1 |
| 0 | 1 | 0 | 0 |
| 0 | 1 | 1 | 1 |
| 1 | 0 | 0 | 0 |
| 1 | 0 | 1 | 1 |
| 1 | 1 | 0 | 0 |
| 1 | 1 | 1 | 1 |

**Table 2**

| $\dfrac{X_2 X_3}{X_1}$ | 00 | 01 | 11 | 10 |
|---|---|---|---|---|
| 0 | 0 | 1 | 1 | 0 |
| 1 | 0 | 1 | 1 | 0 |

### 2.2 Canalizing Function for $n$ variable
Canalizing function as defined in [3] is a type of Boolean function in which at least one of the input variables is able to determine the output of the function regardless of the input values of the other variables. For $n$ variables $(x_1, x_2, ....x_n)$ the degree of the variables ranges from 1 to $n$. In $x_i$ of degree $i$, first $2^{i-1}$ consecutive bits are similar and the next $2^{i-1}$ bits will be complement of the first $2^{i-1}$ bits. This process will continue until all the $2^n$ bits are obtained. If $x_i$ is the canalizing input for a function $f$ then with respect to at least one of the inputs (0 or 1) in $x_i$ the output of the function $f$ will be restricted.

### Example 1: Illustration of Canalizing Function
If $n = 2$, there are 16 Boolean functions as shown in Table 3.All the canalizing functions of 2 variables generated from different canalizing inputs can be obtained from Table 4. '*' marked position in Table 4 can be filled up in four ways {00,01,10,11}. So the total number of distinct canalizing functions obtained for 2 variable are {1100, 1101, 1110, 1111, 0000, 0001, 0010, 0011, 0111, 1011, 0100, 0101, 1000, 1010}

**TABLE 3: All Boolean Functions for 3 Variables**

| $X_1$ | $X_2$ | $f_0$ | $f_1$ | $f_2$ | $f_3$ | $f_4$ | $f_5$ |
|---|---|---|---|---|---|---|---|
| 0 | 0 | 0 | 1 | 0 | 1 | 0 | 1 |
| 0 | 1 | 0 | 0 | 1 | 1 | 0 | 0 |
| 1 | 0 | 0 | 0 | 0 | 0 | 1 | 1 |
| 1 | 1 | 0 | 0 | 0 | 0 | 0 | 0 |

| $X_1$ | $X_2$ | $f_6$ | $f_7$ | $f_8$ | $f_9$ | $f_{10}$ | $f_{11}$ |
|---|---|---|---|---|---|---|---|
| 0 | 0 | 0 | 1 | 0 | 1 | 0 | 1 |
| 0 | 1 | 1 | 1 | 0 | 0 | 1 | 1 |
| 1 | 0 | 1 | 1 | 0 | 0 | 0 | 0 |
| 1 | 1 | 0 | 0 | 1 | 1 | 1 | 1 |

| $X_1$ | $X_2$ | $f_{12}$ | $f_{13}$ | $f_{14}$ | $f_{15}$ |
|---|---|---|---|---|---|
| 0 | 0 | 0 | 1 | 0 | 1 |
| 0 | 1 | 0 | 0 | 1 | 1 |
| 1 | 0 | 1 | 1 | 1 | 1 |
| 1 | 1 | 1 | 1 | 1 | 1 |

**Table 4(Structure for different canalizing inputs)**

| $X_1$ | $X_2$ | Case A | Case B | Case C | Case D |
|---|---|---|---|---|---|
| 0 | 0 | * | * | 1 | 0 |
| 0 | 1 | * | * | 1 | 0 |
| 1 | 0 | 1 | 0 | * | * |
| 1 | 1 | 1 | 0 | * | * |

| $X_1$ | $X_2$ | Case E | Case F | Case G | Case H |
|---|---|---|---|---|---|
| 0 | 0 | 1 | 0 | * | * |
| 0 | 1 | * | * | 1 | 0 |
| 1 | 0 | 1 | 0 | * | * |
| 1 | 1 | * | * | 1 | 0 |

### 2.3 Identification of canalizing boolean function using K-Map
**Theorem1:** Any $n$ variable boolean function can be identified as canalizing boolean function if the entries in at least $x/2$ rows or $y/2$ columns in $K_0(x, y)$ is all 0's or all 1's and either of the following two conditions are satisfied

(i) All entries in $K_1(x/2, y)$ or $K_1^*(x/2, y)$ or $K_1(x, y/2)$ or $K_1^*(x, y/2)$ all 0's or all 1's

(ii) $\sum_{i=1}^{\log_2(x)-1} K_i(x/2^i, y) \sim K_i^*(x/2^i, y)$ holds $\forall i$ or

$\sum_{i=1}^{\log_2(y)-1} K_i(x, y/2^i) \sim K_i^*(x, y/2^i)$ holds $\forall i$

Where $x = 2^{\lceil n/2 \rceil}, y = 2^{\lfloor n/2 \rfloor}$

$K_0(x, y)$ = Karnaugh Map representation of the function $f$,

$K_i(x/2^i, y) = 1^{st}$ $x/2$ rows and $y$ columns of $K_{i-1}(x/2^{i-1}, y)$,

$K_i^*(x/2^i, y)$ = Last $x/2$ rows $y$ columns of $K_{i-1}(x/2^{i-1}, y)$ taken in reverse order

$K_i(x, y/2^i) = 1^{st}$ $x$ rows and $y/2$ columns of $K_{i-1}(x, y/2^{i-1})$

$K_i^*(x, y/2^i) =$ Last $x$ rows and $y/2$ columns of $K_{i-1}(x, y/2^{i-1})$ taken in reverse order.

$K_i(a,b) \sim K_i^*(a,b)$. If in at least $a/2$ no of rows $b$ consecutive columns in both matrices have same value (all 0's or all 1's) and same location or in at least $b/2$ no of columns $a$ consecutive rows in both matrices have same value (all 0's or all 1's) and same location

**Proof:** The above theorem has been proved by the method of induction.

**Basis:** For $n = 2$, 16 boolean functions are present and in the K-Map representation of the boolean functions in 2 variables the number of rows =2 and columns =2. From the K-Map representation of the boolean 1,2,6,9 as seen in Fig1 it can be concluded that for function 1,2 in $K_0(x, y)$ all the entries of $K_1(x/2, y) = 0$, so theorem 1 is satisfied for function 1,2 but for function 6,9 theorem 1 is not satisfied. Similarly by drawing the K-Map for other boolean functions in 2 variable it can be seen that theorem 1 is satisfied. So for $n = 2$ in 14 boolean functions theorem 1 is satisfied. Hence for $n = 2$ theorem 1 is satisfied.

**Hypothesis:** Let us assume that theorem 1 is true for $n = m$. All the boolean functions of $m$ variable canalized with respect to any input $m_i$ will satisfy Theorem 1. So at least $x/2$, rows or $y/2$ columns in $K_0(x, y)$ is all 0's or all 1's and any of the following two conditions is satisfied

i) All entries in $K_1(x/2, y)$ or $K_1^*(x/2, y)$ or $K_1(x, y/2)$ or $K_1^*(x, y/2)$ all 0's or all 1's

(ii) $\sum_{i=1}^{\log_2(x)-1} K_i(x/2^i, y) \sim K_i^*(x/2^i, y)$ or

$\sum_{i=1}^{\log_2(y)-1} K_i(x, y/2^i) \sim K_i^*(x, y/2^i)$

| Table 1: K-Map for f(1) | | |
|---|---|---|
| $x_1/x_2$ | 0 | 1 |
| 0 | 1 | 0 |
| 1 | 0 | 0 |

| Table 2: K-Map for f(2) | | |
|---|---|---|
| $x_1/x_2$ | 0 | 1 |
| 0 | 0 | 1 |
| 1 | 0 | 0 |

| Table 3: K-Map for f(6) | | |
|---|---|---|
| $x_1/x_2$ | 0 | 1 |
| 0 | 0 | 1 |
| 1 | 1 | 0 |

| Table 4: K-Map for f(9) | | |
|---|---|---|
| $x_1/x_2$ | 0 | 1 |
| 0 | 1 | 0 |
| 1 | 0 | 1 |

**FIG 1**

**Induction:** It has to be shown that for $n = m+1$ theorem 1 is true. Now if for $m+1$ variable the function $f$ is canalized with respect to $x_1$, then

$K_0(x, y)$ Can be any of the form shown below

$$\begin{pmatrix} * & * & \cdots & * & * \\ * & * & \cdots & * & * \\ * & * & \cdots & * & * \\ * & * & \cdots & * & * \\ 0/1 & 0/1 & \cdots & 0/1 & 0/1 \\ 0/1 & 0/1 & \cdots & 0/1 & 0/1 \\ 0/1 & 0/1 & \cdots & 0/1 & 0/1 \\ 0/1 & 0/1 & \cdots & 0/1 & 0/1 \end{pmatrix} \begin{matrix} \Big\} 2^m \\ \\ \\ \Big\} 2^m \end{matrix} \begin{pmatrix} * & * & \cdots & * & * \\ * & * & \cdots & * & * \\ * & * & \cdots & * & * \\ * & * & \cdots & * & * \\ 0/1 & 0/1 & \cdots & 0/1 & 0/1 \\ 0/1 & 0/1 & \cdots & 0/1 & 0/1 \\ 0/1 & 0/1 & \cdots & 0/1 & 0/1 \\ 0/1 & 0/1 & \cdots & 0/1 & 0/1 \end{pmatrix}$$

FIG 2

K-Map becomes as in FIG 2. So $K_1(x/2, y)$ or $K_1^*(x/2, y)$ have all entries same. Hence by Theorem 1 if a function $f$ in $m+1$ variable is canalized with respect to $x_1$ then it can be detected by the help of K-Map. Now if a function is canalized with respect to $x_2$ then the K-Map representation is like FIG 3. Considering $K_1(x/2, y)$ and $K_1^*(x/2, y)$ in FIG 3 it is observed that $K_1(x/2, y) \sim K_1^*(x/2, y)$ and also $K_2(x/4, y)$ in $m+1$ variable is same as the K-Map representation of a function $g$ in $m$ variable when canalized with respect to $x_1$. Since by hypothesis, for $m$ variable when a function is canalized with respect to $x_1$ theorem 1 is true so in $m+1$ variable also when $f$ is canalized with respect to $x_2$ theorem 1 is true. Similarly K-Map representation of $f$ canalized with respect to $x_3$ in $m+1$ variable is same as the K-Map representation of a function $g$ in $m$ variable when canalized with respect to $x_2$. From here it can be concluded that K-Map representation of $f$ canalized with respect to $x_i$ in $m+1$ variable will be same as K-Map representation of a function $g$ canalized with respect to $x_{i-1}$ in $m$ variable. Since by hypothesis, for $m$ variable when a function is canalized with respect to any $x_i$ theorem 1 is true so in $m+1$ variable also when $f$ is canalized with respect to any $x_i$ theorem 1 is true. Hence by the principle of mathematical induction it can be concluded that Theorem 1 is true for natural number $m$

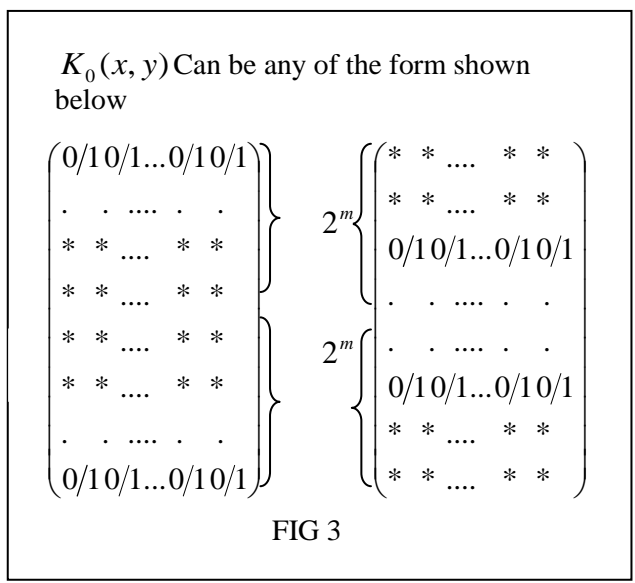

FIG 3

(* marked position of FIG 3 indicates all entries in these locations are either all 0's or all 1's)

## 2.4 Some illustration with examples

*Example 2:* To check whether $f$ = 11010000 1111 00001111000011110000 is a canalizing function or not.
*Solution:* The function $f$ is a 5 variable boolean function canalized with respect to $x_3$. To represent it on K-Map the number of rows and columns should be $2^{\lceil 5/2 \rceil} = 8$ and $2^{\lfloor 5/2 \rfloor} = 4$ respectively. The K-Map representation of the function $f$ i.e. $K_0(x,y)$, $K_1(x/2, y)$, $K_1^*(x/2, y)$, $K_2(x/4, y)$, $K_2^*(x/4, y)$ is shown in FIG4. From FIG 4 it can be seen that there exists 4 rows having all the bits in each column same (all zeros) in $K_0(x,y)$ and $K_1(x/2, y) \sim K_1^*(x/2, y)$, $K_2(x/4, y) \sim K_2^*(x/4, y)$. So by theorem 1 the identification has been done.

*Example 3:* To check whether $f$ = 0000111000 01111111000011110000 is a canalizing function or not
Solution: As can be seen from FIG 5 $K_1(x/2, y) !\sim K_1^*(x/2, y)$ so $f$ is not a canalizing function.

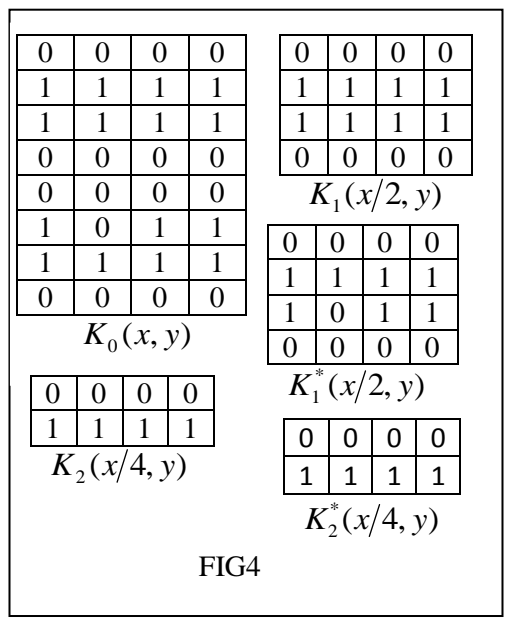

FIG4

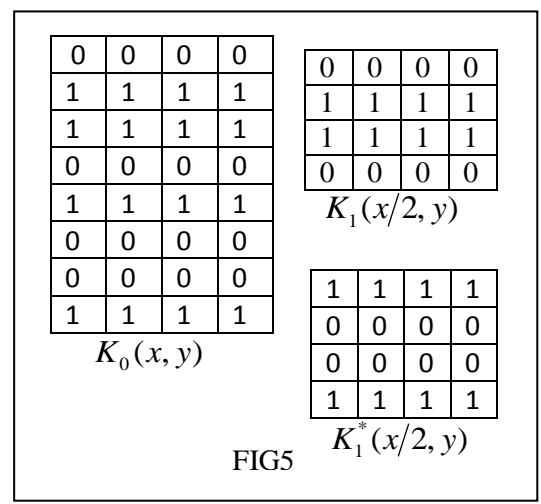

FIG5

## 3. Properties Of Canalizing Functions From Theorem 1.
*Lemma 1:* If $f$ is a canalizing function on $n$ variable then $f$ complement ($f'$) will also be a canalizing function on $n$ variable

***Proof:*** If $f$ is a canalyzing function on $n$ variable then it will satisfy the conditions of theorem 1. Since $f$ is a canalyzing function $f'$ will be just the converse of $f$, the 0's in $K_0(x,y)$ will be replaced by 1's and the 1's by 0's as a result the conditions of theorem 1 which were satisfied by $f$ will also be satisfied by $f'$ now hence $f'$ also becomes a canalyzing function.

***Lemma 2:*** If $f$ is a canalyzing function on $n$ variable then $ff$ is also a canalyzing function on $n+1$ variable.

***Proof:*** Since $f$ is a canalyzing function on $n$ variable theorem 1 is satisfied and let $K_0(x,y)$ be the K-Map representation of $f$ on $n$ variable. When $f$ is concatenated with $f$ then on representing $ff$ on K-Map of $n+1$ variable it is observed that $K_1(x/2,y) \sim K_1^*(x/2,y)$ and $K_1(x/2,y)$ of $ff$ on $n+1$ is same as $K_0(x,y)$ of $f$ on $n$ variable. As $K_0(x,y)$ all the conditions of theorem 1 and $K_1(x/2,y) \sim K_1^*(x/2,y)$, so $ff$ is a canalyzing function on $n+1$ variable.

***Lemma 3:*** If $f$ and $f'$ is a canalyzing function on $n$ variable then $ff'$ or $f'f$ will be a canalyzing function on $n+1$ variable only if $f=0$ or $f=2^{2^n}-1$

***Proof:*** When $f=0$ then $f'=2^{2^n}-1$ and in $n+1$ variable concatenation of $f$ with $f'$ will generate the boolean function $2^{2^{(n+1)-1}}-1$. Representation of $ff'$ in the K-Map is as shown in Fig 6. It is observed that all entries in $K_1(x/2,y)$ is all 0's, so by theorem 1 $ff'$ is a canalyzing function and by lemma1 $f'f$ will also be a canalyzing function.

| 1 | 1 | 1 | 1 |
|---|---|---|---|
| 1 | 1 | 1 | 1 |
| 1 | 1 | 1 | 1 |
| 1 | 1 | 1 | 1 |
| 0 | 0 | 0 | 0 |
| 0 | 0 | 0 | 0 |
| 0 | 0 | 0 | 0 |
| 0 | 0 | 0 | 0 |

$f = 2^{2^n}-1$

$f = 0$

FIG 6

***Lemma 4:*** Each non canalyzing function $f$, concated with an $n$ variable boolean function $g$ can generate only two canalyzing function in $n+1$ variable.

***Proof:*** After concatenation of $f$ with $g$, $fg$ will be a canalyzing function in $n+1$ variable if all the $2^n$ bits of $g$ is either all 0 or all 1. So only two possibilities are present here as shown in FIG 7. If the '*' marked position of $K_0(x,y)$ is either all 0's or all 1's then in both these cases $K_1(x/2,y)$ have all the entries similar, hence by theorem 1 $fg$ will be a canalyzing function.

| 0/1 | 0/1 | 0/1 | 0/1 |
|---|---|---|---|
| 0/1 | 0/1 | 0/1 | 0/1 |
| 0/1 | 0/1 | 0/1 | 0/1 |
| 0/1 | 0/1 | 0/1 | 0/1 |
| * | * | * | * |
| * | * | * | * |
| * | * | * | * |
| * | * | * | * |

$2^n$

$2^n$

FIG 7

***Lemma 5:*** If $f,g$ are two boolean functions in $n$ variable then concatenation of $f,g$ ($fg$) will generate $2^n$ canalizing functions in $n+1$ variable if $f = 0$ or $2^{2^n}-1$

***Proof:*** When $f$ in $n$ variable is concatenated with $g$ then in $n+1$ variable in $K_1(x/2,y)$ all the entries are either all zeros or all ones. So by theorem 1 $fg$ will be a canalyzing function.

***Lemma 6:*** If $f,g$ are two boolean functions in $n$ variable then concatenation of $f,g$ ($fg$) will generate $2 \times \sum_{x=1}^{n} {}^nC_x \times 2^{2^n/2^x}(-1)^{(x-1)}$ number of canalyzing functions in $n+1$ variable if hamming distance (H.D) of $f$ from either 0 or $2^{2^n}-1$ is 1

***Proof:*** If a function $f$ has H.D 1, it is a canalyzing function in $n$ variable, then to become a canalyzing function in $(n+1)$ variable $2^n/2$ bits will be left and the remaining $2^{n+1}-2^{n-1}$ bit position will be occupied. The $2^{n-1}$ spaces can be filled up in $2^{2^n/2}$ ways. A function can be canalized with respect to more than 1 input simultaneously, so these canalyzing functions have to be calculated only once. So each time ${}^nC_x \times 2^{2^n/2^x}(-1)^{(x-1)}$ number of canalizing function has to be eliminated where $x$ is the number of inputs with respect to which the function $f$ in $n+1$ variable is canalized. Hence the number of canalizing function becomes $\sum_{x=1}^{n} {}^nC_x \times 2^{2^n/2^x}(-1)^{(x-1)}$. Now with respect to $f$ concatenation can be done in both left hand side of $f$ as well as right hand side of $f$. So the total number of canalizing functions generated by any canalizing function of $n$ variable having hamming distance 1 is $2 \times \sum_{x=1}^{n} {}^nC_x \times 2^{2^n/2^x}(-1)^{(x-1)}$.

***Lemma 7:*** Concatenation of two non canalizing functions in $n$ variable is non canalizing function in $(n+1)$ variable.

***Proof:*** Any non canalizing function in $n$ variable will be a canalizing function in $(n+1)$ if lemma 4 is satisfied. So concatenation of two non canalizing functions becomes non canalizing.

***Lemma 8:*** If $X$ is the number of canalizing and in $n$ variable then all the canalizing functions in $n+1$ variables can be generated by using $(X-2)^2 - (X-2)$ number of concatenation operations in $n$ variable.

***Proof:*** There are $2^{2^n}$ boolean functions and $X$ canalizing functions in $n$ variable. Then the number of non canalizing functions in $n$ variable is $2^{2^n} - X$.

i) By lemma 7 concatenation of two non canalizing functions is non canalizing. So the number of non canalizing functions generated in $(n+1)$ variable from $2^{2^n} - X$ non canalizing functions in $n$ variable are
$(2^{2^n} - X) \times (2^{2^n} - X)$ .................. (1)

ii) By lemma 4 any non canalizing function in $n$ variable can generate 2 canalizing functions in $n+1$ variables. So total number of canalizing functions generated is
$2 \times (2^{2^n} - X)$ .......... (2)

iii) By lemma 5 it follows that canalizing functions $f = 0$ or $2^{2^n} - 1$ when concatenated with any non canalizing function will also generate a canalizing function in $(n+1)$. So from here the number of canalizing functions generated is $2 \times (2^{2^n} - X)$ ...... (3)

iv) Concatenation of non canalizing functions with canalizing functions other than $f = 0$ or $2^{2^n} - 1$ will form a non canalizing function in $n+1$ variable (follows from lemma 4). Total number of such occurrences is
$(2^{2^n} - X) \times (X - 2) \times 2$ ...... (4)

v) Concatenation of two canalizing functions in $n$ variable can generate canalizing as well as non canalizing functions in $n+1$ variables. Total possible outcomes here is
$X \times X$ ...... (5)

To identify all the canalizing functions in $n+1$ variable Theorem1 had to be checked $2^{2^{n+1}}$ times. But by the method of concatenation it has been observed that (1) and (4) always generates non canalizing function and (2),(3) will be always generating canalizing function in $(n+1)$ variable, hence they need not to be checked using K-Map.

Again by lemma 5 concatenation of any canalizing function with $f = 0$ or $2^{2^n} - 1$ will always be a canalizing function, so eliminating these two functions concatenation and identification has to be done on $(X-2)$ number of canalizing functions. By lemma 2, lemma 3 concatenation of any function $f$ with $f$ is canalizing, and concatenation of $f$ with $f'$ is non canalizing, so $(X-2)$ number of concatenation can be eliminated.

So by using $(X-2)^2 - (X-2)$ concatenation operations in $n$ variable all the canalizing functions in $n+1$ variable can be identified.

## 4. NESTED CANALYZING FUNCTION

Nested canalyzing function (NCF) are special type of canalyzing function and they are defined in [6, 15] which states that if $f$ be a boolean function in $n$ variable and $\sigma$ be a permutation on $\{1,2,...n\}$ then the function $f$ is NCF in the variable order $x_{\sigma 1}, x_{\sigma 2},....x_{\sigma n}$ with canalyzing input values $a_1, a_2,...a_n$ and canalized output values $b_1, b_2,....b_n$ respectively, if it can be represented in the form

$$f(x_1, x_2,...x_n) = \begin{cases} b_1 \text{ if } x_{\sigma(1)} = a_1 \\ b_2 \text{ if } x_{\sigma(2)} = a_2, x_{\sigma(1)} \neq a_1 \\ ............ \\ b_n \text{ if } x_{\sigma(n)} = a_n, x_{\sigma(1)} \neq a_1,...\text{and } x_{\sigma(n-1)} \neq a_{n-1} \\ \neg b_n \text{ if } x_{\sigma(1)} \neq a_1 ...\text{and } x_{\sigma(n)} \neq a_n \end{cases}$$

The function $f$ is nested canalyzing if $f$ is canalyzing in the variable order $x_{\sigma 1}, x_{\sigma 2},....x_{\sigma n}$ for some permutation $\sigma$.

**Hamming Distance (H.D) of Nested Canalyzing Function (NCF):** H.D of N.C.F $f$ from this section has been defined as $\min\{H.D(f,0), H.D(f,2^{2^n}-1)\}$

**NOTE:-** If $x_1, x_2, \ldots, x_n$ are the input variables and more than one permutation order $(\sigma)$ exists for a particular NCF, then the order taken for the NCF is in increasing order of the input variables $x_i$.

## 4.1 MERGER OPERATION IN NCF

In case of $n$ variable boolean functions each of the variables $x_1, x_2, \ldots x_i, \ldots, x_n$ has $2^n$ bits and degree of each of the variables ranges from 1 to $n$. For a function $f$ in $n$ variable $2^n$ bits are present and when $f$ is merged with $x_i$, $2^n$ more bits are added as a result $f$ after merging with $x_i$ becomes a function in $n+1$ variable. Let this function be termed as $g$. The merging operation is done as follows:-

If $x_1$ is merged with $f$ then four outcomes are possible as stated below:-

1) Case1: In the left side of $f$, $2^n$ 0's are merged i.e. $g = (0000\ldots 0000)f$
2) Case2: In the left side of $f$, $2^n$ 1's are merged i.e. $g = (1111\ldots 1111)f$
3) Case3: In the right side of $f$, $2^n$ 0's are merged i.e. $g = f(0000\ldots 0000)$
4) Case 4: In the right side of $f$, $2^n$ 1's are merged i.e. $g = f(1111\ldots 1111)$

From above it can be concluded that when $x_i$ is merged with $f$ then either in the beginning 1st $2^{n-(i-1)}$ bits will be either all 0 or all 1, the next $2^{n-(i-1)}$ bits will be taken from $f$ and the next $2^{n-(i-1)}$ bits will be same as the 1st $2^{n-(i-1)}$ bits. This process continues unless all the $2^{n+1}$ bits obtained.

*Lemma 9:* For $n$ variable boolean function the total number of NCF generated having a particular H.D $j$ with starting canalizing input $x_i$ can be given by the formula

$$N_c = 4 \times \sum_{i=1}^{n} \sum_{j=1}^{2^{n-2}} M_n[i][j] \text{ where } n > 2 \text{ and}$$

$$M_n[i][j]$$
$$= 0; \text{if } i = n \ \& \ 1 \leq j \leq 2^{n-3}$$
$$= M_n[1][2^{n-2}+1-j] \text{ if } 2^{n-3} < j \leq 2^{n-2}$$
$$= 2 \times \sum_{i=i_1}^{n-1} M_{n-1}[i][j] \text{ if } 0 < j \leq 2^{n-3} \ \& \ i \leq i_1 < n$$

$$M_2[i] = \binom{2}{0},$$

where $(i,j) \to (\text{row, column})$

The columns denote the minimum hamming distance. The value $M_n[i][j] = p$ denotes that in case of $n$ variable boolean function there are $p$ NCF having H.D $2 \times j - 1$ from 0 or $2^{2^n}-1$ if starting canalizing input is $x_i$

**PROOF:** The above lemma has been proved by the method of mathematical induction.

**BASIS:** There are 64 NCF in 3 variable boolean functions. From the above formula also it will be shown that 64 NCF are present in case of 3 variable boolean functions. For $n = 3$ the number of rows will be 3 and the number of columns will be 2.
The cell values at the different locations of the matrix $M_3[3][2]$ has been calculated from the above formula and the result obtained is as follows

$$M_3[1][1] = 2 \times \{M_2[1][1] + M_2[2][1]\}$$
$$= 2 \times (2+0) = 4$$
$$M_3[2][1] = 2 \times M_2[2][1] = 0; M_3[3][1] = 0;$$
$$M_3[1][2] = M_3[1][2^{3-2}+1-2] = M_3[1][1] = 4$$
$$M_3[1][2] = M_3[2][2] = M_3[3][2] = M_3[1][1]$$
$$M_3[1][1] = 4$$

Now the matrix $M_3[3][2] = \begin{bmatrix} 4 & 4 \\ 0 & 4 \\ 0 & 4 \end{bmatrix}$

The no of NCF for 3 variables is given as
$$N_c = 4 \times \sum_{i=1}^{3} \sum_{j=1}^{2^{3-2}} M_3[i][j] = 4 \times (4+4+4+4) = 64.$$ Hence the lemma is valid for $n = 3$.

**HYPOTHESIS:** Let us assume that the theorem is valid for $n = m$. So the number of rows and columns will be $m$ and $2^{m-2}$ respectively. Possible hamming distance for $m$ variable is $1,3,5,7,\ldots 2^{m-1}-1$. The matrix $M_m[i][j]$ has been shown below.

$$M_m[i][j] = \begin{pmatrix} P_{11} & .. & 2\times(m-1)\times P_{1,x} & 2\times(m-1)\times P_{1,x} & .. & 2\times P_{11} \\ 0 & .. & 2\times(m-2)\times P_{1,x} & 2\times(m-1)\times P_{1,x} & .. & 2\times P_{11} \\ 0 & .. & 2\times(m-3)\times P_{1,x} & 2\times(m-1)\times P_{1,x} & .. & 2\times P_{11} \\ . & .. & . & . & .. & 2\times P_{11} \\ . & .. & . & . & .. & 2\times P_{11} \\ . & .. & 2\times P_{1,x} & 2\times(m-1)\times P_{1,x} & .. & 2\times P_{11} \\ 0 & .. & 0 & 2\times(m-1)\times P_{1,x} & .. & 2\times P_{11} \end{pmatrix}$$

The number of NCF for $m$ variables can be calculated S $N_c = 4 \times \sum_{i=1}^{m} \sum_{j=1}^{2^{m-2}} M_m[i][j]$

**INDUCTION:** It has to be shown that for $m+1$ variable the formula becomes $N_c = 4 \times \sum_{i=1}^{m+1} \sum_{j=1}^{2^{m+1-2}} M_{m+1}[i][j]$.

Now $2^m$ bits are present in a $m$ variable boolean function. To get a boolean function in $m+1$ variable $2^m$ more bits are merged using merger operation. In case of $m$ variable the different hamming distances are $1,3,5,7....$ $2^{m-1} - 1$. If all 0's ($2^m$ 0's) are merged with those functions having hamming distance 1 from the boolean function 0 in $m$ variable, then boolean functions having hamming distance 1 in $m+1$ variable will be generated, and if all 1's are merged then hamming distance will be $2^m - 1$.

Similarly functions having hamming distance 3 in $m$ variable will generate functions of hamming distance $3, 2^{m-1} - 3$ in $m+1$ variable by the application of merger operation. So for $m+1$ variables possible hamming distance generated are $1, 3, 5, 7, ..... 2^m - 1$. In case of $m+1$ variable no of columns will be $2^m/2 = 2^{m-1} = 2^{m+1-2}$ and no of rows will be $m+1$. Each of $P_{ij}$ in $M_m[i][j]$ can be given by $P_{ij} = \begin{cases} 2\times(m-1)\times M_{m-1}[i][j] \text{ if } j > 2^{m-3} \\ 2\times \sum_{i=i}^{m-1} M_{m-1}[i][j] \text{ if } j \leq 2^{m-3} \end{cases}$

From $M_m[i][j]$ the values at each of the cell locations in $M_{m+1}[i][j]$ can be calculated and the matrix obtained is

$$M_{m+1}[i][j] = \begin{pmatrix} 2\times P_{11} & .. & 2\times\sum_{i=1}^{m-1} M_m[i][j] & 2\times\sum_{i=1}^{m-1} M_m[i][j] & .. & 2\times P_{11} \\ 0 & .. & 2\times\sum_{i=2}^{m-1} M_m[i][j] & 2\times\sum_{i=1}^{m-1} M_m[i][j] & .. & 2\times P_{11} \\ 0 & .. & 2\times\sum_{i=i}^{m-1} M_m[i][j] & 2\times\sum_{i=1}^{m-1} M_m[i][j] & .. & 2\times P_{11} \\ . & .. & . & . & .. & 2\times P_{11} \\ . & .. & . & . & .. & 2\times P_{11} \\ . & .. & 2\times M_m[i][j] & 2\times\sum_{i=1}^{m-1} M_m[i][j] & .. & 2\times P_{11} \\ 0 & .. & 0 & 2\times\sum_{i=1}^{m-1} M_m[i][j] & .. & 2\times P_{11} \end{pmatrix}$$

Each of the cell locations in the above matrix $M_{m+1}[i][j]$ can be viewed as

$P_{ij} = \begin{cases} 2\times(m)\times M_m[i][j] \text{ if } j > 2^{m-2} \\ 2\times \sum_{i=i}^{m-1} M_m[i][j] \text{ if } j \leq 2^{m-2} \end{cases}$

Or $P_{ij}$ can be written as

$P_{ij} = \begin{cases} 2\times(m+1-1)\times M_{m+1-1}[i][j] \text{ if } j > 2^{m+1-3} \\ 2\times \sum_{i=i}^{m+1-2} M_{m+1-1}[i][j] \text{ if } j \leq 2^{m+1-3} \end{cases}$

By principle of mathematical induction it can be seen that the matrix $M_{m+1}[i][j]$ is valid for any natural number $m$. As seen from the merger operation four possibilities are present whenever one function is merged with any of the input variables, hence the number of nested canalyzing function formed from $m+1$ variable is $4\times \sum_{i=1}^{m+1} \sum_{j=1}^{2^{m+1-2}} M_{m+1}[i][j]$. Hence the formula is valid for $m+1$ So by principle of mathematical induction it can be seen that the formula is valid for all natural number $n$.

**ILLUSTRATION WITH EXAMPLE:**

For n=4 the number of NCF is 736.

$M_3[3][2] = \begin{pmatrix} 4 & 4 \\ 0 & 4 \\ 0 & 4 \end{pmatrix}$

The size of the matrix for 4 variable will be $4\times 2^{4-2} = 4\times 4$, the different hamming distance for 4 variable NCF is 1,3,5,7.

The matrix $M_4[4][4] = \begin{pmatrix} 8 & 24 & 24 & 8 \\ 0 & 16 & 24 & 8 \\ 0 & 8 & 24 & 8 \\ 0 & 0 & 24 & 8 \end{pmatrix}$

Number of NCF= $N_c = 4 \times \sum_{i=1}^{m} \sum_{j=1}^{2^{m-2}} M_m[i][j]$

$4 \times (8+ 24+ 16 +8+24+24+24+24+8+8+8+8)$
=736
From the formula given in lemma 9 also the same result is obtained.

## 5. PARTIALLY NESTED CANALYZING FUNCTIONS (P.N.C.F)

The definition of P.N.C.F can be obtained from [6] which states that if $f$ be a boolean function in $n$ variables and suppose for a permutation $\sigma$ on $S_n$, some depth $d \in N, 0 < d < n$ and a boolean function $g(x_{\sigma(d+1)}, x_{\sigma(d+2)}....x_{\sigma(n)})$

$$f(x_1, x_2,...x_n) = \begin{cases} b_1 \text{ if } x_{\sigma(1)} = a_1 \\ b_2 \text{ if } x_{\sigma(2)} = a_2, x_{\sigma(1)} \neq a_1 \\ ............ \\ b_n \text{ if } x_{\sigma(d)} = a_d, x_{\sigma(1)} \neq a_1,... x_{\sigma(d-1)} \neq a_{d-1} \\ g \text{ if } x_{\sigma(1)} \neq a_1 ...\text{and} \quad x_{\sigma(d)} \neq a_d \end{cases}$$

Where either $g$ is a constant function or a non canalyzing function in $(n-1)$ variable.

Here an attempt has been taken to detect the number of P.N.C.F possible of different depths for 4 variables. For 4 variables the different depths of P.N.C.F are 1,2,3.

### 5.1 P.N.C.F of depth 1 for 4 variables
When there are P.N.C.F of depth 1 for 4 variables then the function $g$ can be a constant function or a non canalyzing function in 3 variables.

**CASE 1: When the function $g$ is a constant function.**
Let
$f$ = Set of variables $\{x_1, x_2, x_3, x_4\}$
Y = Set of P.N.C.F of depth 1
$g$ = Constant function ($g$ will be complement of $f$)
Y will be of the form $fg'$ or $gf'$
$|Y| = |gf'| + |fg'| = 8$

0 and 65535 are also P.N.C.F of depth 1.
So $|Y| = 10$

**CASE 2: When the function $g$ is a non canalyzing function**

Let
X= Set of non canalyzing function in 3 variables $|X| = 136$
$f$ = Set of variables $\{x_1, x_2, x_3, x_4\}$
Y = Set of P.N.C.F of depth 1
Y can be of the form $fg$ or $gf$ or $fg'$ or $gf'$
So $|Y| = 4 \times |X| \times |f| = 4 \times 136 \times 4 = 2176$

### 5.2 P.N.C.F of depth 2 for 4 variables

**Case 1: When the function $g$ is a constant function.**

The depth of P.N.C.Fs considered here is 2; so out of 4 variables any 2 variables will act as a canalyzing input. Hence out of 16 bits output for 12 bits will be obtained from the canalyzing input and the remaining 4 bits will be a constant function

Let the order $\sigma = (x_{\sigma_p}, x_{\sigma_q}); p < q$
Y = Set of P.N.C.F of depth 2
$f$ = Output obtained from canalyzing i/p $x_{\sigma_q}$
$h$ = Output obtained from canalyzing i/p $x_{\sigma_p}$
$g$ = Constant function.

Out of 4 variables 2 variables can be selected in $^4C_2 = 6$ ways. So there are 6 different orders. Y can be any of the following forms

$Y = [hfg$ or $hgf$ or $h'fg$ or $h'gf$ or $gfh$ or $fg'h$ or $gfh']$
So for any particular order there exists 8 possible outcomes, hence for 6 different orders total $6 \times 8 = 48$ P.N.C.F of depth 2 will be generated.

**CASE 2: When the function $g$ is a non canalyzing function**
The permutation order of $\sigma$ will contain 2 elements. Just like case 1 here also output of the 12 bits will be obtained from the permutation order $\sigma = (x_{\sigma_p}, x_{\sigma_q})$, the remaining 4 bits will be a non canalyzing function in 2 variables. Considering permutation order $\sigma = (x_{\sigma_p}, x_{\sigma_q})$ $p < q$, $^4C_2$ types of different orderings are possible. Let
$g$ = Set of non canalyzing function.
$f$ = Output obtained from canalyzing i/p $x_{\sigma_q}$
$h$ = Output obtained from canalyzing i/p $x_{\sigma_p}$
Y = Set of non canalyzing function of depth 2

$|Y| = {}^4C_2 \times g \times [hfg \parallel hgf \parallel h'gf \parallel h'gf \parallel fgh \parallel gfh \parallel fgh' \parallel fh'g \parallel hgf' \parallel h'f'g \parallel h'gf' \parallel f'gh \parallel gf'h \parallel f'gh' \parallel gfh'$

$= {}^4C_2 \times 2 \times 16 = 192$

## CASE 3: Considering permutation order $\sigma = (x_{\sigma_q}, x_{\sigma_p})$; $p < q$

Here the total P.N.C.F's are
${}^4C_2 \times g \times [hfg \parallel hgf \parallel h'fg \parallel h'gf \parallel fgh \parallel gfh \parallel fgh' \parallel gfh$
$= 6 \times 2 \times 8 = 96$

### 5.3 P.N.C.Fs of Depth 3 for 4 variables

For P.N.C.Fs of 4 variables if depth is 3; then the function $g$ will be a constant function. Here out of 16 bits, output for 14 bits is obtained from the input canalized values and remaining 2 bits will be a constant function.

Let $\sigma$ be the permutation for the P.N.C.Fs of depth 3 where $\sigma = (x_{\sigma_p}, x_{\sigma_q}, x_{\sigma_r})$ where $p < q < r$

Y=Set of P.N.C.Fs of depth 3 for 4 variable

$f$ = Output obtained from canalizing i/p $x_{\sigma_p}$

$h$ = Output obtained from canalizing i/p $x_{\sigma_q}$

$e$ = Output obtained from canalizing i/p $x_{\sigma_r}$

g=constant function

**Case 1:** $\sigma = (x_{\sigma_p}, x_{\sigma_q}, x_{\sigma_r})$ where $p < q < r$

Y will be the form,
$|Y| = hfeg \parallel hegf \parallel h'feg \parallel h'egf \parallel fegh \parallel egfh \parallel fegh' \parallel egfh' \parallel$ (replacing $f$ & $e$ with their complements in these 8 combinations)

So $|Y| = {}^4C_1 \times 8 \times 2 \times 2 = 128$

**Case 2:** $\sigma = (x_{\sigma_q}, x_{\sigma_p}, x_{\sigma_r}) \parallel (x_{\sigma_q}, x_{\sigma_r}, x_{\sigma_p})$

where $p < q < r$

Y will be the form,
$|Y| = [fegh \parallel fheg \parallel fgeh \parallel fhe'g' \parallel f'egh' \parallel f'h'eg \parallel f'geh \parallel f'he'g' \parallel egfh \parallel hfeg \parallel gefh \parallel hfeg \parallel egf'h' \parallel f'h'eg \parallel f'h'ge \parallel gef'h']$

So $|Y| = {}^4C_1 \times 16 = 64$

**Case 3:** $\sigma = (x_{\sigma_r}, x_{\sigma_p}, x_{\sigma_q}) \parallel (x_{\sigma_r}, x_{\sigma_q}, x_{\sigma_p})$

where $p < q < r$

Y will be the form,

$|Y| = egfh \parallel efgh \parallel ehfg \parallel ehgf \parallel e'g'fh' \parallel e'fgh' \parallel e'h'gf \parallel e'h'fg \parallel gefh \parallel fegh \parallel hegf \parallel hefg \parallel ge'fh' \parallel fe'gh \parallel he'fg \parallel he'gf$

So $|Y| = {}^4C_1 \times 16 = 64$

Total number of partially nested canalizing functions of depth 3 = (128+ 64+64) =256. Now total number of canalyzing functions for 4 variables are 3514. By adding the total number of P.N.C.F and N.C.F of different depths the result obtained is (2186+336+256+736) which is equal to 3514.

## 6. CONCLUSION AND DISCUSSION

Through the use of K-Map it can be easily checked whether any arbitrary $n$ variable boolean function is canalizing or not. It can be concluded that if all the canalizing functions in $n$ variable is known, then by the method of concatenation all the canalizing functions in $n+1$ variable can be generated very easily. Through one of the formulas given in this paper, the number of N.C.F having a particular H.D with any arbitrary Boolean variable as starting canalizing input can be detected.

For 4 variable Boolean functions the number of partially nested canalizing functions having different depths has been calculated in this paper. Our further aim is to generalize this concept and detect the number of partially nested canalizing functions having different depth for any $n$ variable Boolean functions.

The further aim is to study different gene regulatory networks with the help of interaction graph and nested canalizing functions.